\documentclass[fleqn,twoside]{article}
\usepackage[headings]{espcrc2}

\readRCS
$Id: espcrc2.tex,v 1.2 2004/02/24 11:22:11 spepping Exp $
\usepackage{graphicx, balance}
\usepackage[figuresright]{rotating}

\usepackage{epstopdf}

\usepackage{cite}
\usepackage[cmex10]{amsmath}
\usepackage{enumerate} 
\usepackage[english]{babel}
\usepackage{algorithm}
\usepackage[noend]{algpseudocode}
\usepackage{multirow}
\usepackage{makecell}
\usepackage{array,booktabs,calc}
\usepackage{enumitem}

\title{\textbf{Performance Analysis of RFDMRP:River Formation Dynamics based Multi-Hop Routing Protocol in WSNs}}

\author{Koppala Guravaiah\address[DCSE]{Department of Computer Science and Engineering, National Institute of Technology, Tiruchirappalli 620015, India, Contact: kguravaiah@gmail.com \\},
R. Leela Velusamy\address[DCSE]{Department of Computer Science and Engineering, National Institute of Technology, Tiruchirappalli 620015, India, Contact: leela@nitt.edu}}

\runtitle{Performance Analysis of RFDMRP}
\runauthor{Koppala Guravaiah and  R. Leela Velusamy}

\begin{document}
\begin{abstract}

In Wireless sensor networks, sensor nodes sense the data from environment according to its functionality and forwards to its base station. This process is called Data collection. The Data collection process is done either directly or by multi-hop routing. In direct routing, every sensor node directly transfers its sensed data to base station which has an impact on energy consumption from sensor node due to the far distance between the sensor node and base station. In multi-hop routing, the sensed data is relayed through multiple nodes to the base station, it consumes less energy. This paper presents and analyzes the performance of a data collection routing protocol called RFDMRP: River Formation Dynamics based multi-hop routing protocol. The performance of RFDMRP is tested and analyzed for network parameters such as Network lifetime, Energy usage, and Node density \& data aggregation impact on network lifetime. The simulated results are compared with two algorithms LEACH and MOD\_LEACH. The comparison reveals that the proposed algorithm performs better than LEACH and MOD\_LEACH with respect to Network lifetime.  \\\\
{\bf Keywords :} Data Collection, Energy efficiency, Network lifetime, River Formation Dynamics. 

\end{abstract}

\maketitle

\section{INTRODUCTION}
Wireless Sensor Networks\cite{Akyildiz2002survey,rault2014energy,yick2008wireless} (WSNs) are widely used in various real time applications such as military, medical, disaster detection,  structural monitoring, etc. These WSNs consists of huge set of small sensor nodes, deployed in the environment for monitoring environmental conditions such as humidity, temperature, pressure, etc. The wireless sensor nodes sense the data from environment based on the application and forwards to the central base station or sink for further processing\cite{wang2011networked}. This process is called data collection, which is the primary task of the WSNs. In data collection process\cite{wang2011networked}, the sensor nodes forward the data to the central base station either by direct communication or by multi-hop communication. The direct communication from sensor node to base station is energy expensive due the distance between sensor nodes and base station is more, this reduces the lifetime of the network. Alternatively, Multi-hop communication\cite{stavrou2010survey,al2004routing,pantazis2013energy} schemes are used for better network lifetime and performance due to its effective utilization of resources. In multi-hop communication, every sensor node is busy in forwarding the sensed/received data to nearest intermediate (neighbor) nodes or to the base station using multi-hop routing paths. In this process, selection of next (neighbor) node in routing path is very important for forwarding data. The next node or forwarding node in the routing path is not only meant for relaying the data, but also useful for aggregating the data. Data aggregation or Data fusion techniques are used to reduce the size of the data packet to be transmitted to next node by aggregating the data or by eliminating similar data, received from previous nodes\cite{rault2014energy}. Multi-hop techniques improve the energy conservation of node and the lifetime of the network.

Swarm intelligence is one of the mechanism used for finding the suitable nodes in the routing path between sensor nodes and the base station. In WSNs, swarm intelligence mechanisms\cite{zungeru2012classical,lin2012energy,amin2014smart} such as ant colony, bee colony, etc., are already used to select the next node in the routing path. A nature inspired mechanism known as River Formation Dynamics (RFD)\cite{rabanal2007using} can be introduced in WSNs for suitable node selection in multi-hop routing. RFD mechanism is free from local cycles and this is one of the facts making it suitable for path finding in WSNs. The RFDMRP (River formation Dynamics based Multi-Hop Routing Protocol) was proposed for WSNs by exploring the applicability of RFD mechanism in path finding\cite{Guravaiah2015RFDMRP}. The two parameters hop count distance and residual energy are used by RFD for selecting the suitable nodes. In this paper, the performance of RFDMRP is analyzed and compared with the existing algorithm such as LEACH and MOD\_LEACH. The Node density, data aggregation, network lifetime, and energy consumption are used for comparison.

Henceforth the paper is organized as follows: Section \ref{Sec:Back} discusses the conceptual RFD and radio energy model. The related work is explained in Section \ref{Sec:related}. Section \ref{Sec:problem} describes the problem statement. The RFDMRP data collection protocol implementation details is discussed in Section \ref{Sec:rfdmrp}. The simulation results are analyzed and compared in Section \ref{Sec:simulation}. Finally, Section \ref{Sec:conclusion} concludes the paper.

\section{BACKGROUND} \label{Sec:Back}
\subsection{River Formation Dynamics (RFD)}

RFD\cite{rabanal2007using} is one of the heuristic optimization method and a subset topic of swarm intelligence. RFD is based on replicating the concept of how water drops combine to form rivers and rivers in turn combine to join the Sea by selecting the shortest path based on altitudes of the land through which they flow. In the process of river formation, the water drops are always flowing from higher altitude position to lower altitude positions. Since, the slope of the two positions is more, then the water flowing from higher positions to lower positions erode and carry the eroded soil to be deposited in the lower positions. By this deposit the altitude of the lower position get increased. Also shortest path is formed from higher to lower position. 

\begin{algorithm}
\caption{General RFD algorithm}
\label{alg:RFD}
\begin{algorithmic}
\Procedure{RFD\_Algorithm}{}
\Statex \textbf{\hspace{0.5 cm}//Stage I: Initialization Stage}
\State \text{Initialization of Drops generating positions;}
\State \text{Initialization of Intermediate positions;}
\State \text{Initialization of Destination(Sea) positions;}
\Statex \textbf{\hspace{0.5 cm}//Stage II: River Formation Stage}
\While{ {(not all drops Flow The Same Path) and (not other Ending Condition)}}
  \State ${select\_Forward\_Position();}$
	\State ${move\_Drops();}$     
	\State ${erode\_Path();}$
	\State ${add\_Sediments();}$
 \EndWhile
\State \textbf{end} 
	\State \text{Analyze the paths;}

\EndProcedure
\State \textbf{end procedure}
\end{algorithmic}
\end{algorithm}

The basic algorithm of RFD is given in Algorithm \ref{alg:RFD}. This algorithm mainly consist of two stages viz., Initialization stage and River formation stage. In initialization stage, three different positions (called water drop generating positions or Source(S), intermediate positions(I), and destination(D)  or sea) are initialized. All these positions are represented with different  altitude values (S and I are represented with positive altitude values and D is represented with Zero). The water drop generating positions always generates water drops. The intermediate positions receives the water drops from source and forward towards the Sea. In river formation stage, the river is created between drop generating positions and Sea using the iterative process having the functions ${select\_Forward\_Position()}$, ${move\_Drops()}$, ${erode\_Path()}$, and ${add\_Sediments()}$. The iterative process is repeated until either all drops follow the same path or satisfying the other ending conditions such as limited number of iterations, limited execution time. In ${select\_Forward\_Position()}$, the drop generating positions select the next neighbor positions for forwarding the drops based on the probability function $P(i,j)$ in Equation \ref{Eq:prob}, where i, j are the positions such that ($i \epsilon S$ or $i \epsilon I$) and ($j \epsilon I$ or $j \epsilon D$).  The probability function $P(i,j)$ indicates that a position $i$ having the probability to select the position $j$ as a next hop position for forwarding drops.
\begin{equation}
            \label{Eq:prob} 
						P(i,j) =            
						\begin{cases}
                     \frac{DG(i,j)}{\sum_{l \epsilon Nb(i)}DG(i,l)}	 &  if j \epsilon Nb(i)\\
                     0	 &  Otherwise
                    
						\end{cases}
\end{equation}
            where $Nb(i)$ is the neighbor positions of position $i$ and
where $DG(i,j)$ is Decreasing Gradient between node $i$ and node $j$ and it can be calculated using the following Equation \ref{Eq:dg}
\begin{equation}
\label{Eq:dg}
        DG(i,j) = \frac{(altitude(i)-altitude(j))}{distance(i,j)}
 \end{equation}

In the function $erode\_Path()$, according to drop movements the paths are eroded. If a drop moves from position
A to position B then we erode A and deposit that eroded soil to B using function $add\_Sediments()$. That is, the altitude of A position is reduced and altitude value of B position is increased
depending on the current gradient between A and B. If the down slope between A and B is high then the erosion
is higher. The altitude of the destination position (Sea) is never modified and it remains equal to 0 during all the execution. Finally, analyze the paths formed by drops and stores the optimized path.

There is a similarity between RFD and data collection processes in WSN is given in Table \ref{Tab:rfd_wsn_diff}. In RFD, the source (drop generating) positions generate water drops and these water drops are interested to meet the destination or Sea. Similarly, in WSN data collection process, the sensor nodes generate the data and this data is interested to reach the base station. Hence, the sensor data act like water drops, the source positions like sensor nodes, and base station as Sea. The drops are combined and flows from source to sea to form the rivers based on altitude value of position in RFD. In the same way the sensor nodes can form a path to the base station for forwarding data in WSNs based on hop-count and residual energy. 
		\begin{table}
		\caption{similarity between RFD and data collection processes in WSN}
			\label{Tab:rfd_wsn_diff}
			\begin{tabular}{|p{0.8in}|p{1.6in}|} \hline
            {\textbf{RFD}}&\textbf{WSN} \\ \hline

             Water Drops             &   Sensor nodes          \\ \hline
             Sea                     &   Base Station \\ \hline
             River                   &   Path    \\ \hline    
             Altitude                &   Parameters such as Hop Count, Residual Energy, etc.    \\ \hline    
      \end{tabular}
		\end{table}
\subsection{Energy Model}
The first order radio energy model of sensor node considered for this work is discussed in literature                          \cite{heinzelman2002application}. In this energy model, the sensor node consumes $E_{TX-elec}$ (Transmitter electronics) energy to run transmitter circuit and consumes $E_{RX-elec}$ (Receiver electronics) energy to run receiver circuit. If a sensor node wants to transfer $'b'$ bit of data in $d$ distance, then the energy consumption at transmitter $E_{TX} (b,d)$ is calculated as in Equation \ref{Eq:trans}. 
\begin{equation}
\label{Eq:trans}
     E_{TX} (b,d) =        
     \begin{cases}
                     b ( E_{TX-elec} +  \epsilon_{fs} d^2 )	 &  d \le d_0\\
                     b ( E_{TX-elec} +  \epsilon_{mp} d^4 )	 &  d > d_0
                    
     \end{cases}                
 \end{equation}

where $\epsilon_{fs}$ is energy required by the transmitter amplifier in the free space model. $\epsilon_{mp}$ is the energy needed by the transmitter amplifier in multi-path model and $d_0$ is the Threshold value and it is calculated using, $\sqrt(\frac{\epsilon_{fs}}{\epsilon_{mp}})$.
Similarly, the energy consumption for receiving $'b'$ bit of data, $(E_{RX} (b))$, is calculated using Equation \ref{Eq:rec}  
\begin{equation} 
\label{Eq:rec}
E_{RX} (b)= b E_{RX-elec}   						
\end{equation} 
If a node wants to perform data aggregation or data fusion on $'b'$ bit data packet, then the energy consumption $(E_{F} (b))$ for data aggregation at that node is given in Equation \ref{Eq:Aggregation}  
\begin{equation} 
\label{Eq:Aggregation}
E_F (b)= b  E_{DA}						
\end{equation} 
where $E_{DA}$ is the energy needed for Data Aggregation or fusion of sensor data. 
\section{RELATED WORK} \label{Sec:related}
During the last decade, researchers have extensively investigated various techniques on multi-hop routing. The existing multi-hop routing techniques are mainly based on hierarchical or cluster based routing protocols\cite{heinzelman2002application,heinzelman2000energy,mahmood2013modleach,lindsey2002pegasis,farooq2010mr,biradar2011multihop,ma2013improvement}.

Heinzelman et al.\cite{heinzelman2000energy} proposed LEACH (Low Energy Adaptive Clustering Hierarchy), a hierarchical cluster based routing algorithm for enhancing the network lifetime and reducing the energy usage. In this protocol, the network is divided into clusters, which consist of a set of cluster members that are managed by a cluster head. The cluster member sends the data to respective cluster head and cluster heads forward this data to base station. Cluster heads are selected randomly in a distributed manner. Later, LEACH was modified to LEACH-C\cite{heinzelman2002application}, where the cluster head selection process was based on a centralized process, i.e., cluster heads are elected by the base station based on their residual energy. Mahmood et al.\cite{mahmood2013modleach} proposed an MODLEACH, modified version of LEACH. In MODLEACH, a cluster head replacement technique and dual transmitting power levels were proposed to improve the performance based on the metrics such as throughput and lifetime. However, in above techniques due to dynamic cluster formation the distance between Cluster Head (CH) and the Base Station (BS) is far away and some of the cluster node are also far away from its cluster heads. Hence, direct data transmission from cluster members to CH and CH to BS leads to more energy consumption. Later, S. Lindsey et al.\cite{lindsey2002pegasis} enhanced the LEACH protocol and proposed PEGASIS, a multi-hop chain-based protocol. A chain is formed by sensor nodes in such a way that every node participates in transmitting and/or receiving  data from a neighbor node. One node is selected from the chain for forwarding the collected data in chain to the BS. In the chain, the collected sensor data are aggregated and carried from node to node and finally the same are transferred to the BS. PEGASIS provides better performance than LEACH by reducing the cluster formation overhead, minimizing the number of transmissions between sensor nodes and BS. However, data transmission delay is more due to the large chain length in PEGASIS. In literature \cite{ma2013improvement,biradar2011multihop,farooq2010mr}, enhanced LEACH protocols are proposed. In these protocols, the multi-hop communication is introduced in between the CH and BS to improve the lifetime of the network.  

\section{PROBLEM STATEMENT} \label{Sec:problem}
A Graph $G(V,E)$ represents the WSN, where $V$ is the set of $n$ number of Nodes( Sensor Nodes and a Base station $S$) and $E$ is the set of  wireless links between nodes. In each Round, all sensor nodes $V_i$, $\forall i=1,2,..n$ in network forward its data to Base station $S$. The problem is to maximize the number of rounds before the network dies. This is achieved by identifying the energy efficient multi-hop paths $P_E(V_i \rightarrow S)$ from each sensor node ($V_i$) to the Base station $S$. The main aim is to select the next hop node for finding the energy efficient multi-hop paths from each sensor node ($V_i$) to $S$. In this paper, RFD mechanism is used to find the next hop node for establishing an energy efficient paths.

\section{RFDMRP: RFD BASED MULTI-HOP ROUTING PROTOCOL} \label{Sec:rfdmrp}
\begin{center}
  \begin{figure}[!t]
    \includegraphics[width=2.9in]{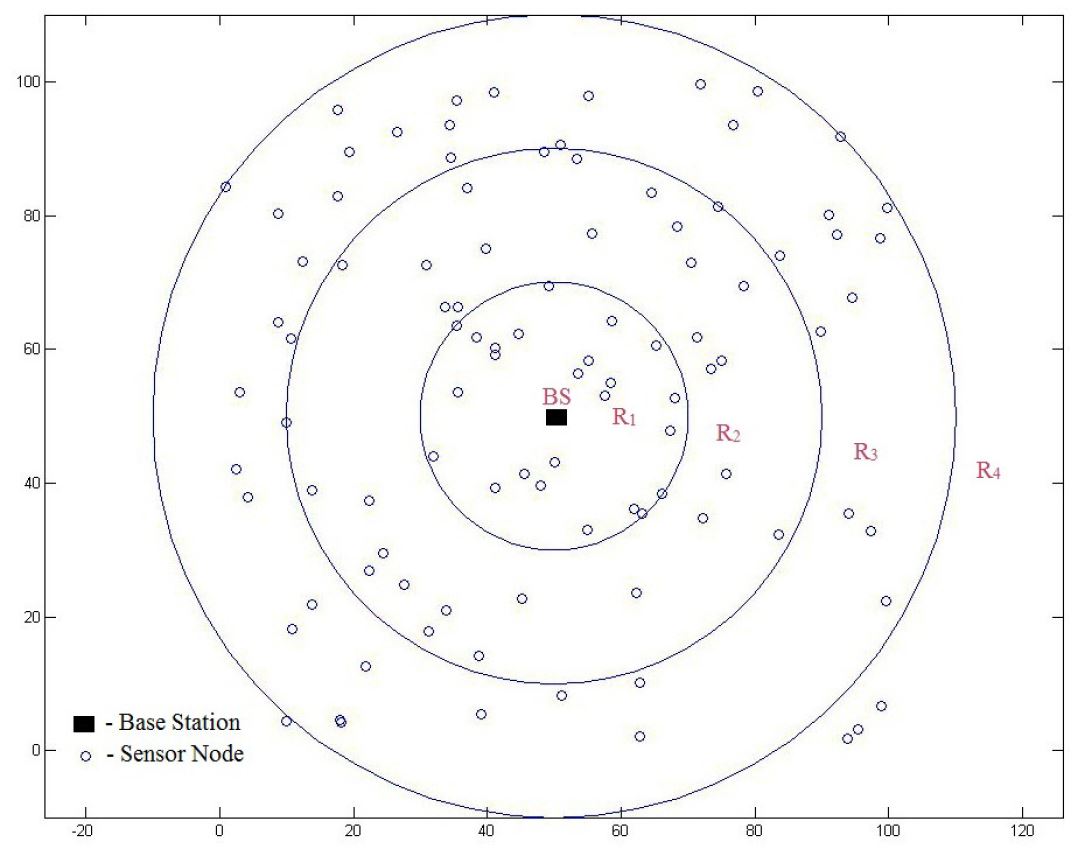}
    \caption{Example Network Scenario with various regions.}
    \label{fig:Scenario}
  \end{figure}
\end{center}

RFDMRP, a multi-hop routing protocol, is proposed for data collection in WSN. An example network is given in Figure \ref{fig:Scenario} having 100 randomly deployed sensor nodes and a base station presented in middle of the network. From the Figure \ref{fig:Scenario}, The base station divides the network into various regions such as $R_1$, $R_2$, $R_3$, etc., using the following Equation \ref{Eq:division}. 

\begin{equation}
\label{Eq:division}
     \text {Number of Regions} = \frac {max\_dist}{T_r/2}                
 \end{equation}
where 
 $max\_dist = max (distance(BS, V_1),$ $distance(BS, V_2),$ $...,distance(BS, V_i))$  $\forall i \epsilon V$ , 
 $T_r$ is transmission range of sensor nodes and $distance(BS, V_i)$ is the Euclidean distance between Base Station and Sensor Node $V_i$  (i.e., $\sqrt{(x_{BS}-x_{V_i})^2 + (y_{BS}-y_{V_i})^2 }$). Where $(x_{BS}, y_{BS})$ and $(x_{V_i}, y_{V_i})$ are the coordinates of Base Station and sensor node($V_i$) respectively.

The proposed RFDMRP algorithm is explained in Algorithm \ref{alg:RFDMRP}. This algorithm consists of two stages: Initialization stage and Path Selection and Data Relay stage. These stages are explained in the section given below.   

\begin{algorithm}
\caption{RFDMRP algorithm}
\label{alg:RFDMRP}
\begin{algorithmic}
\Procedure{$RFDMRP\_Algorithm()$}{}
\Statex \textbf{\hspace{0.5 cm}//Stage I: Initialization Stage}
\State ${nodeDeployment()}$
\State ${NNTableCreation()}$
\Statex \textbf{\hspace{0.5 cm}//Stage II: Path Selection and Data Relay}
\While{ \text{(not all nodes are dead) }}
  \Repeat
  \State ${forward\_Node\_Selection()}$
  \State $forward\_Data()$
  \State $update\_Energy()$
  \State $update\_NNtable()$
  \Until{\text{data reaches to BS}}
\EndWhile
\State \textbf{end}

\EndProcedure
\State \textbf{end procedure}
\end{algorithmic}
\end{algorithm}

\subsection{Initialization Stage}
\begin{figure}[!t]
\centering
\includegraphics[width=2.9 in]{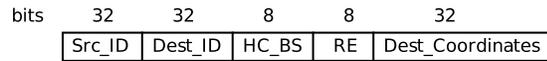}
\caption{REQUEST/REPLY Packet Format.}
\label{fig:req}
\end{figure}

Initially, in this stage, all the sensor nodes are deployed in the environment based on the application. All nodes in the network calculate its hop count distance from the BS. For calculation of hop count, BS broadcasts the  Beacon message containing its identity. The node, which receives the Beacon signal responds with its id and its location coordinates. BS calculate the hop count from each node based on the node coordinates and send the hop count information to nodes.
Each sensor node stores hop count value in Neighbor Node information table (NN table) shown in Table \ref{NNtable}. The NN table consist of Next\_Node, Hop Count between Next node and BS (HC\_BS), Neighbor Node Remaining Energy (RE(NN)), Distance (Distance between source node and next node), and Distance from next node to BS (D\_to\_BS). To calculate the neighbor node information, source nodes (Src\_ID) sends a REQUEST packet to the neighboring nodes. The neighboring (Dest\_ID) node upon receiving REQUEST packet, search in its NN table for HC\_BS, NNRE, and Coordinates. Then, it replies with the REPLY packet to the source node (Src\_ID) then source node updates its NNtable. The format of REQUEST and REPLY packets are shown in Figure \ref{fig:req}.  

\begin{table}[t]
\centering
\caption{Neighbor Node (NN) Table}
\label{NNtable}
\scalebox{0.75}{
\begin{tabular}{|c|c|c|c|c|} \hline
{Next\_Node}&{HC\_BS}&{RE(NN)}&{Distance}&{D\_BS} \\ \hline
{$V_1$}&{4}&{$RE(V_1)$}&{0}&{45.50} \\ \hline
{$V_2$}&{2}&{$RE(V_2)$}&{15.23}&{25.23} \\ \hline
. &{.}&{.}&{.}&{.} \\
. &{.}&{.}&{.}&{.} \\
. &{.}&{.}&{.}&{.} \\ \hline

\end{tabular}}
\end{table}

\begin{figure}[!t]
\centering
\includegraphics[height=0.5in, width=2 in, keepaspectratio]{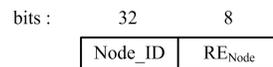}
\caption{ENERGY\_LEVEL Packet Format.}
\label{fig:EnergyLevel}
\end{figure}

\subsection{Path Selection and Data Relay Stage}
In this stage, path is selected between source nodes and the BS by selecting the forward  node using RFD mechanism. Once the path is selected, the source node uses the selected path to relay the data to BS.  This stage consists of three steps: 1) Forward Node Selection 2) Aggregate and Forward the Data 3) Update Energy of each Node and NN information table. 

\subsubsection{Forward Node Selection} 
In the proposed approach source node sends the sensed data to the sink node using multi-hop communication. In multi-hop communication, the selection of neighbor node for onward data forwarding is an important task. Here RFD mechanism is used for next hop node selection. 

Lets assume that the node $i$ is present in Region $R_n$ and node $j$ present in Region $R_m$, where $m \le n$. For forwarding or receiving data, the Residual Energy of both node $i$, and node $j$ must be more than Threshold value($T_E$) i.e., $RE(i) \geq T_E$ and $RE(j) \geq T_E$. The node $i$ having the probability to select node $j$ as a forward node is denoted by $P(i, j)$ and is calculated as follows:

\begin{equation}
		\label{Eq:prob1}
				 P(i,j) =        
				 \begin{cases}
												\frac{H(i, j)}{\sum{H(i, l)}}   & {if (j\epsilon NN_i \& H(i, j)>0)} \\
																	 0                    & Otherwise
												
				 \end{cases}  
 \end{equation} 

where $l\epsilon NN_i$ and  $H(i,l)>0$.

\begin{equation}
\label{Eq:hcd}
     H(i, j)  =  \frac{HC(i,BS)  - HC(j,BS)}{ distance(i,j)}.RE(j)                
 \end{equation}

where
$HC$ is called Hop Count.
$distance(i,j)$ is the euclidean distance between node $i$ to node $j$ (i.e., $\sqrt{(x_{i}-x_{j})^2 + (y_{i}-y_{j})^2 }$ ).
$NN_i$ is Neighbor Node list of node $i$, i.e. $NN_i=\{j,$ such that $distance(i,j) \leq T_r$\}.
$RE(j)$ is residual energy of node $j$. 
$T_E$ is threshold value = 20\% of Initial Energy ($E_0$).

\subsubsection{Aggregate and Forward the Data}
Forward nodes perform data fusion or aggregation on receiving and/or sensed data and then forward to the next selected forward node towards the BS. The energy consumption for data aggregation is calculated using the Equation \ref{Eq:Aggregation}. 
 
\subsubsection{Update Energy of each Node and NN information table}
The energy model explained in Section II is used to calculate the RE (Residual Energy) of a node i, when it transmits or receives the data packet as follows:
  \begin{equation}        
     RE(i) = RE(i)- (E_{TX} (b,d) + E_{RX} (b,d) + E_F(b))         
 \end{equation}
Later, all the nodes updates energy of each node in its NN information table by exchanging the ENERGY\_LEVEL packet. The format of ENERGY\_LEVEL packet is as shown in Figure \ref{fig:EnergyLevel}. 

\begin{figure}[!t]
\centering
\includegraphics[height=3.7in,width=2.9in]{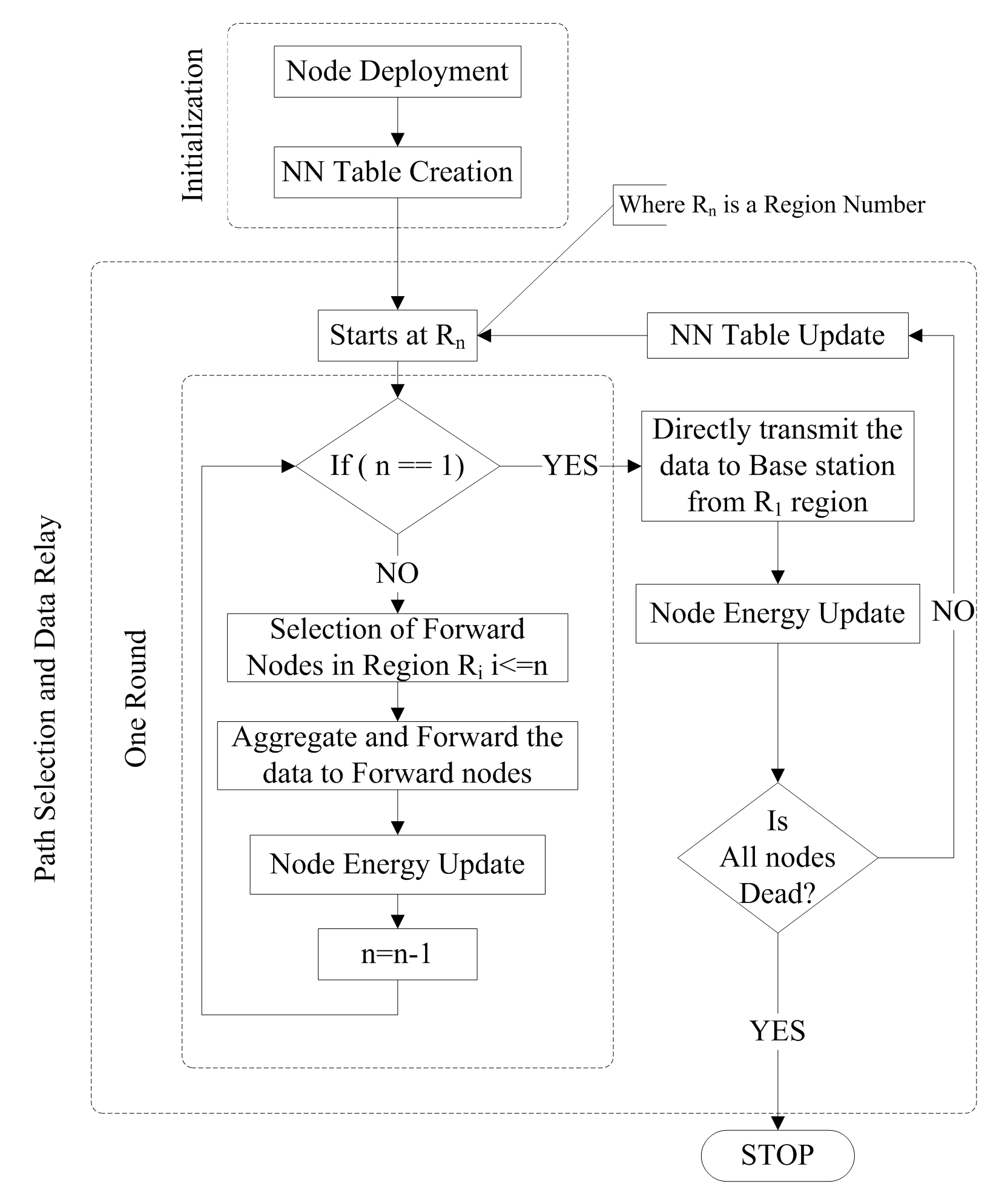}
\caption{Process flow diagram of RFDMRP.}
\label{fig:flow}
\end{figure}

The steps involved in proposed RFDMRP algorithm is given as flowchart in Figure \ref{fig:flow}. This algorithm is executed in the form of rounds. Each round starts at Region $R_n$ and ends at BS. The rounds (process) stop only when all the nodes in the network are dead.
   
\section{SIMULATION RESULTS AND ANALYSIS} \label{Sec:simulation}
The proposed RFDMRP routing algorithm is developed and tested using the MATLAB (2012b). The algorithm is simulated using the simulated parameters listed in Table \ref{table_1}. 

\begin{table}[t]
\centering
\caption{Simulation Parameters}
\label{table_1}
\scalebox{1}{
\begin{tabular}{|l|l|} \hline
{Parameter}&Value \\ \hline

 Number of nodes             &   100          \\ \hline
 Network size                &   100m X 100m \\ \hline
 BS location                 &   (50, 50)    \\ \hline
 Initial node energy($E_o$)  &   0.5 J       \\ \hline
 $E_{elec}$                  &   50 nJ/bit    \\ \hline
 $\epsilon_{fs}$             &   10 pJ/bit/$m^2$ \\ \hline
 $\epsilon_{mp} $            &   0.0013 pJ/bit/$m^4$ \\ \hline
 $E_{DA}$                    &   5 nJ/bit     \\ \hline
 Data packet size            &   4096 bytes  \\ \hline
 Transmission Range          &   20m         \\ \hline
\end{tabular}}
\end{table}

\subsection{Result Analysis}
In this Section the performance of proposed algorithm RFDMRP is analyzed and compared with LEACH and MODLEACH protocols. For comparison important performance parameters are considered such as 
\begin{enumerate}[nolistsep]
  \item \textbf{Alive Nodes:} Nodes which are having more than the energy threshold value ($T_E$) and participate in communication.
	\item \textbf{Dead Nodes:} Nodes which are having less than the energy threshold value ($T_E$) and these nodes will not participate in communication.
	\item \textbf{Packets sent to Base Station:} Total Number of packets transferred to Base Station.
	
	\item	\textbf{Energy Consumption of Network:} Difference between Total Energy and Total Energy Consumed at particular time instance.
	\item \textbf{Network Lifetime:} Time duration from the network initialization to network termination. 
	\item	\textbf{Node Density:} The total number of nodes present in a given network area.
	\item \textbf{Data Aggregation:} It is the process of aggregating the data and reducing the size of the data. Data Aggregation mechanism is applied at each CHs, These CHs collect the data from CMs, aggregating that data, and forwarding to BS. Let us consider a cluster $C$ having k members, then k members sense the b-bit data forward this to CH, CH aggregate the data and forwards to BS. The C is having aggregate mechanism as follows:
	\begin{equation}
				\label{Eq:agg}
							D_A = \gamma * b                                             
	\end{equation}
	$\delta$ is aggregate factor lies in the interval [0, 1]. 
	Where $\gamma=0$ indicates full aggregation i.e., even if any number of packets sent to CH, it will convert into a single packet.
		$\gamma=1$ indicates no aggregation i.e., it will not perform any aggregation.
\end{enumerate}

\begin{figure}[!t]
\centering
\includegraphics[width=2.9 in]{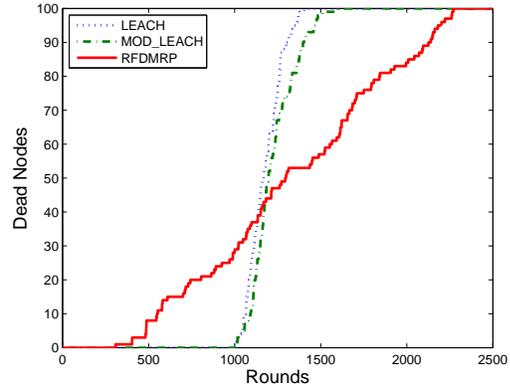}
\caption{Dead nodes Vs Rounds.}
\label{fig:dead}
\end{figure}

\begin{figure}[!t]
\centering
\includegraphics[width=2.9 in]{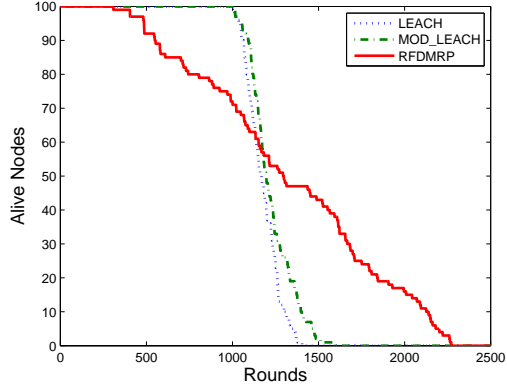}
\caption{Alive nodes Vs Rounds.}
\label{fig:alive}
\end{figure}
Figure \ref{fig:dead} shows the graph plotted for the number of dead nodes over simulation rounds. The first node died earlier in RFDMRP than the other two approaches due to multi-hop transmission of data packets. However, the last node expired earlier in existing approaches than the RFDMRP. This shows the lifetime is extended in the proposed approach due to the dynamic selection of the next hop node.  Similarly, the graph plotted for number of alive nodes over simulation rounds is shown in Figure \ref{fig:alive}.

\begin{figure}[!t]
\centering
\includegraphics[width=2.9 in]{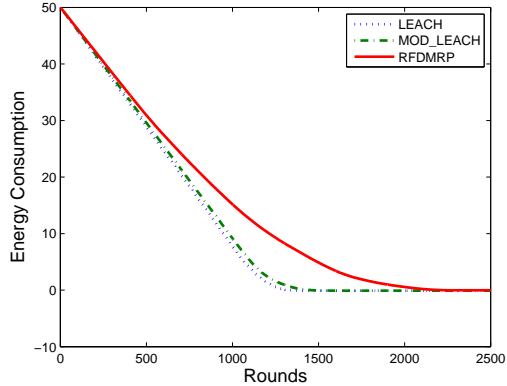}
\caption{Remaining Energy Vs Rounds.}
\label{fig:energy}
\end{figure}

Figure \ref{fig:energy} shows the graph plotted for remaining energy of each round. RFDMRP consumed less energy compared to the existing algorithms. This is due to the nodes in the RFDMRP transfer the data to the nearest (less distance) node with less energy.  

\begin{figure}[!t]
\centering
\includegraphics[width=2.9 in]{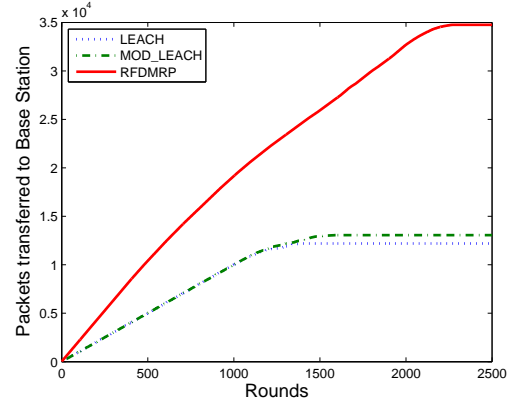}
\caption{Number of data packets transferred to Base Station Vs Rounds.}
\label{fig:bs}
\end{figure}

\begin{figure*}[!t]
\centering
\includegraphics[height=4 in, width=5.5 in ]{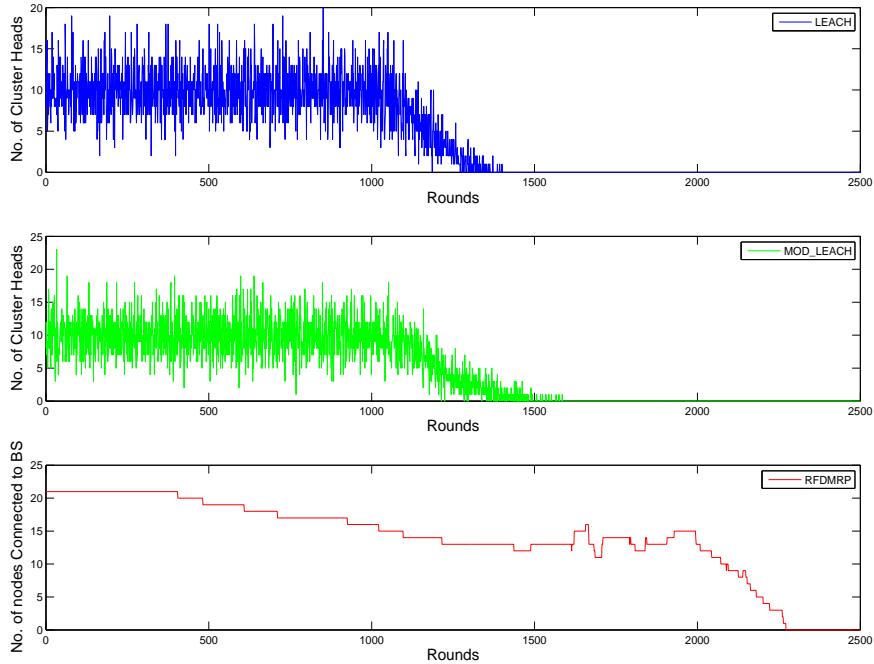}
\caption{Number of Nodes directly communicated with Base Station Vs Rounds.}
\label{fig:node}
\end{figure*}

Figure \ref{fig:bs} shows the graph plotted between data packets transmitted to base station and simulation rounds. More data packets are transmitted to the Base station in RFDMRP when compared to the existing approaches. This is due to the nodes nearer to the BS (in region 1) in RFDMRP are more than the number of cluster heads in existing approaches as shown in Figure \ref{fig:node}.

\begin{figure}[!t]
\centering
\includegraphics[width=2.9 in]{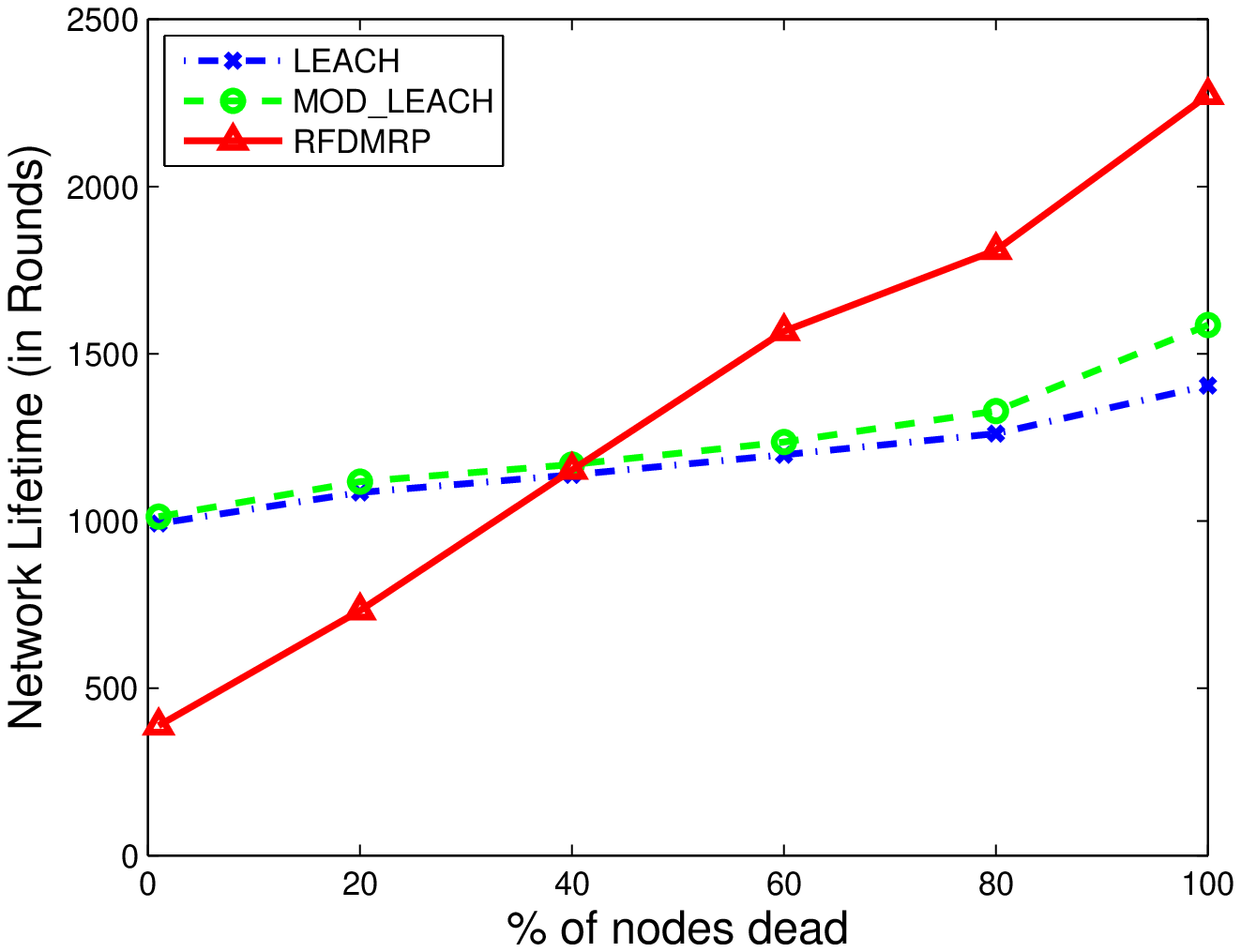}
\caption{Network Lifetime.}
\label{fig:life}
\end{figure}

The network lifetime of proposed and existing protocols is showed in Figure \ref{fig:life}. The nodes which are not having the neighbors are directly connected to BS and these nodes lost their energy quickly.   

\begin{figure}[!t]
\centering
\includegraphics[width=2.9 in]{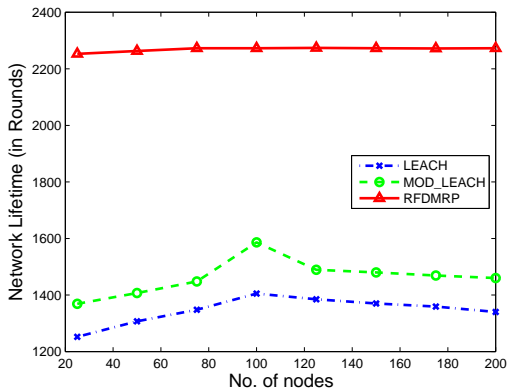}
\caption{Density of Network Vs Network Lifetime.}
\label{fig:density}
\end{figure}

\begin{figure}[!t]
\centering
\includegraphics[width=2.9 in]{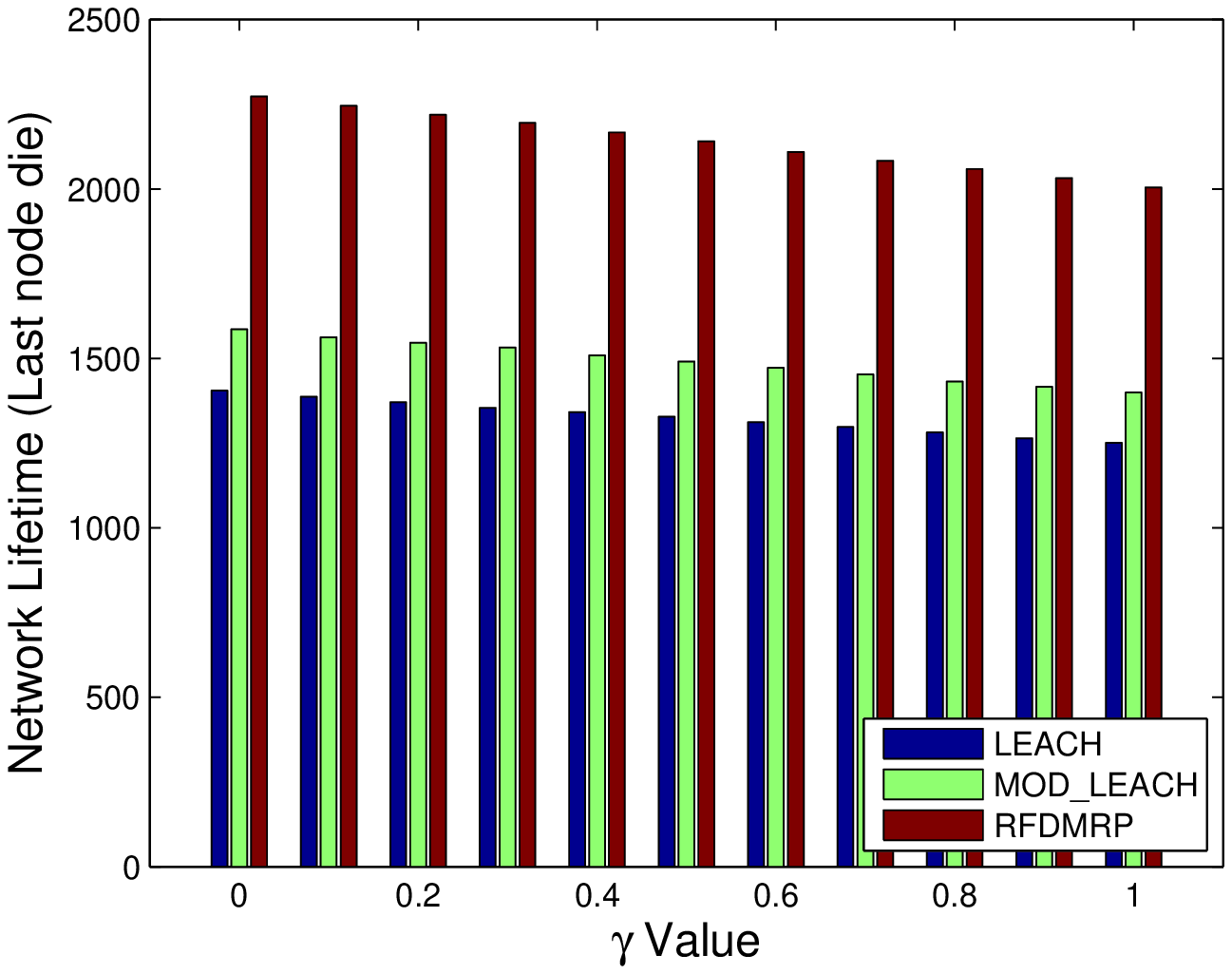}
\caption{Data Aggregation.}
\label{fig:aggregation}
\end{figure}

Figure \ref{fig:density} shows the impact on network lifetime by varying the node density in existing and proposed protocols. The Figure \ref{fig:density} was plotted with 25 nodes to 200 nodes in the $100 \times 100$ network area. There is increase in number of nodes within network which leads to congestion and this affects the network lifetime of existing protocols. Where as in Proposed protocol, the multi-hop mechanism balances the energy consumption and maintain the network lifetime equally in any type of network with any number of nodes.

The lifetime of the network was calculated by changing the $\gamma$ value in the Equation \ref{Eq:agg} and Figure \ref{fig:aggregation} was plotted. In Figure \ref{fig:aggregation}, lifetime of the network was considered as a simulation round when the last node dies. The lifetime is decreasing when the $\gamma$ is increasing from 0 to 1. Here the $\gamma=0$ indicates the full aggregation of data, which results the aggregator takes b-bit data and aggregates that into single data packet. This leads to decreasing in transmission  packets, which in turn reduces the energy consumption for transmission. In the case of $\gamma=1$, The aggregation will not taken place due to this aggregator simply forwards the packets as many as it received. This consumes energy more hence decreasing in network lifetime.

\section{CONCLUSION} \label{Sec:conclusion}
In WSN, multi-hop routing is an effective mechanism for data collection. In multi-hop routing, the selection of forward node for relaying data plays a vital role. One of the swarm intelligence mechanisms, RFD, is used to propose RFDMRP. RFDMRP,  is an RFD based multi-hop routing protocol for data collection in WSN to save energy and enhance the lifetime of the network. In RFDMRP, RFD considers the hop count value and residual energy as parameters for forward node selection. In this paper, the RFDMRP performance was analyzed and compared with LEACH and MOD\_LEACH by considering the performance metrics such as network lifetime and energy consumption, Node density, and data aggregation. From the results, it is observed that RFDMRP performs better than the existing algorithms.

\bibliographystyle{unsrt}
\bibliography{references}

\noindent{\includegraphics[width=1in,height=1.7in,clip,keepaspectratio]{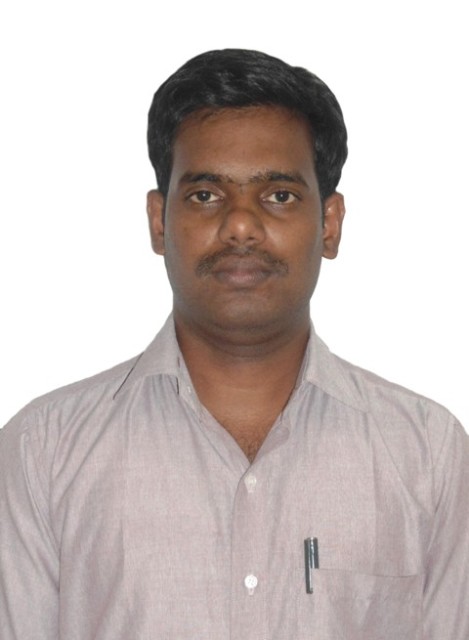}}
\begin{minipage}[b][1in][c]{1.8in}
{\centering{\bf {Koppala Guravaiah}} obtained his degree in Computer Science and Engineering in 2009 from Audisankara College of Engineering and Technology, JNTU Ananthapur and Post graduate degree in Computer Science in 2011 from Sree Vidyanikethan }\\\\
\end{minipage}
Engineering College, JNTU Ananthapur, Andhra Pradesh, India. He is currently doing research in National Institute of Technology (NIT), Tiruchirappalli, Tamil Nadu, India. His research interest includes Data collection routing in Wireless Sensor Networks and Routing in Mobile Ad Hoc Networks. \\\\
\noindent{\includegraphics[width=1in,height=1.7in,clip,keepaspectratio]{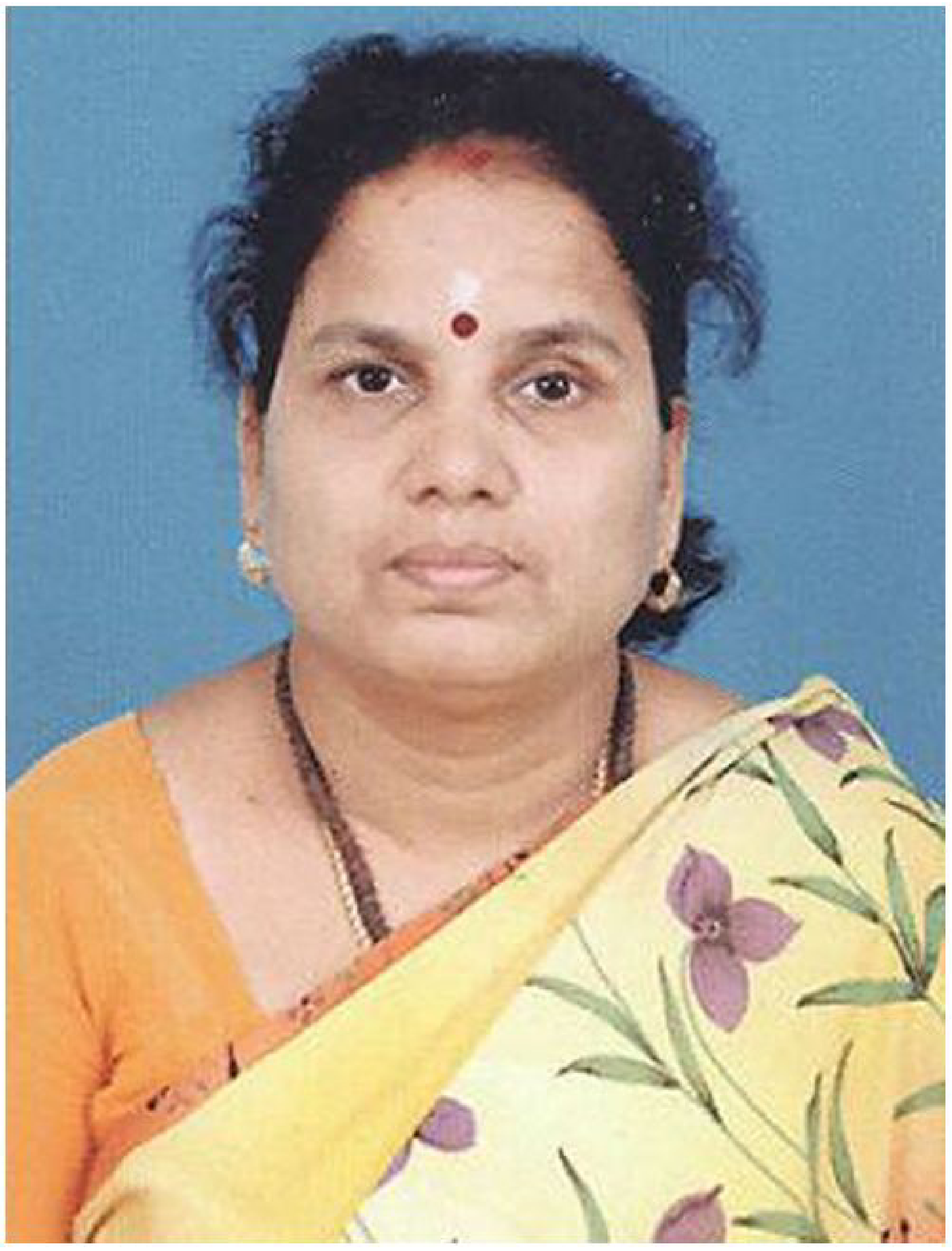}}
\begin{minipage}[b][1in][c]{1.8in}
{\centering{\bf{R. Leela Velusamy}} obtained her degree in Electronics and Communication Engineering in 1986 from REC Tiruchirappalli and Postgraduate degree in Computer science and engineering in 1989 from REC Tiruchirap-}\\\\
\end{minipage}
palli. She was awarded Ph.D. degree by the NIT Tiruchirappalli in 2010. Since 1989, she has been in teaching profession and currently she is a Associate professor in the Department of CSE, NIT Tiruchirpapalli, Tamil Nadu,India. Her research interests include QoS routing, Ad hoc and Wireless Sensor Networks, Social Networks and Digital forensics.\\\\
\end{document}